\documentclass[conference]{IEEEtran}
\IEEEoverridecommandlockouts
\usepackage{amsthm,amsmath,amssymb,mathtools,bm,etoolbox,float,hyperref}
\usepackage[dvipsnames]{xcolor}
\usepackage{graphicx,cite,cuted}
\usepackage{bigints}
\usepackage{caption}

\usepackage{stackengine}

\usepackage{balance}

\theoremstyle{corollary}
\newtheorem{corollary}{Corollary}

\usepackage{mathrsfs,verbatim}
\usepackage{ellipsis}
\usepackage{tikz}
\usepackage{tikz-3dplot}
\usepackage[utf8]{inputenc}
\usepackage{pgfplots} 
\usepackage{pgfgantt}
\usepackage{pdflscape}
\pgfplotsset{compat=newest} 
\pgfplotsset{plot coordinates/math parser=false}
\pgfplotsset{every  tick/.style={black,},ylabel style={font=\tiny},xlabel style={font=\tiny},tick label style={font=\tiny},legend style= {font=\scriptsize},
minor x tick num=1,minor y tick num=1,xminorticks=true,yminorticks=true,}
  \newlength\fheight
\newlength\fwidth

\usepackage{xinttools} 
\usepackage{tikz}
\usepackage{tkz-euclide}
\usetikzlibrary{calc}
\newtheorem{theorem}{Theorem}
\newtheorem{proposition}{Proposition}
\newtheorem{lemma}{Lemma}
\newtheorem{remark}{Remark}

\def\biglen{20cm} 
\tikzset{
  half plane/.style={ to path={
       ($(\tikztostart)!.5!(\tikztotarget)!#1!(\tikztotarget)!\biglen!90:(\tikztotarget)$)
    -- ($(\tikztostart)!.5!(\tikztotarget)!#1!(\tikztotarget)!\biglen!-90:(\tikztotarget)$)
    -- ([turn]0,2*\biglen) -- ([turn]0,2*\biglen) -- cycle}},
  half plane/.default={1pt}
}

\DeclareMathAlphabet{\pazocal}{OMS}{zplm}{m}{n}

\usepackage{caption}
\usepackage{multicol,subcaption}
\usepackage{balance}

\def\BibTeX{{\rm B\kern-.05em{\sc i\kern-.025em b}\kern-.08em
    T\kern-.1667em\lower.7ex\hbox{E}\kern-.125emX}}
\begin{document}

\title{Reverse Link Analysis for Full-Duplex Cellular Networks with Low Resolution ADC/DAC}

\author{\IEEEauthorblockN{Elyes Balti and Brian L. Evans}\thanks{This work was supported by AT\&T Labs and NVIDIA, affiliates of the 6G@UT Research Center within the Wireless Networking and Communications Group at The University of Texas at Austin.}
\IEEEauthorblockA{\textit{6G@UT Research Center} \\
\textit{Wireless Networking and Communications Group} \\
\textit{The University of Texas at Austin} \\
ebalti@utexas.edu, bevans@ece.utexas.edu}
}

\maketitle

\begin{abstract}
In this work, we consider a full-duplex (FD) massive multiple-input multiple-output (MIMO) cellular network with low-resolution analog-to-digital converters (ADCs) and digital-to-analog converter (DACs). Our first contribution is to provide a unified framework for reverse link (uplink) analysis where matched filters are applied at the FD base stations (BSs) under channel hardening. Second, we derive the expressions of the signal-to-quantization-plus-interference-plus-noise ratio (SQINR) for general and special cases. Finally, we quantify effects of quantization error, pilot contamination, and full duplexing for a hexagonal cell lattice on spectral efficiency and cumulative distribution function (CDF) to show that FD outperforms half duplex (HD) in a wide variety of scenarios.
\end{abstract}

\begin{IEEEkeywords}
Full-Duplex, Massive MIMO, Low Resolution Data Converters, Cellular Networks, Interference.
\end{IEEEkeywords}

\section{Introduction}
Full-duplex (FD) has emerged as a viable solution to enable efficient use of spectrum, double spectral efficiency, and reduce latency \cite{ianMagazine}. These benefits are achieved because the FD transceiver transmits and receives in the same resource block in time and frequency \cite{ianTWC}. In addition, an FD transceiver reduces hardware costs by employing a single shared antenna array for both transmission and reception. Thanks to these advantages, FD has been deployed in fielded backhaul systems, prototyped for machine-to-machine communications and integrated access and backhaul systems, and supported in 3GPP release 17 of the cellular standard \cite{release17,elyesiab,R6,R7}. 

Although FD systems deserve attention, they are vulnerable to the loopback self-interference (SI) because of the simultaneous transmission and reception in the same resource block \cite{elyesjoint}. When an FD array transmits a signal, the signal is received by the transmitting FD array with much greater power than that signals from other transmitters. In this near-far problem, the SI signal is several orders of magnitude greater than the desired received signal, which renders the FD systems dysfunctional \cite{zf}. For example, in cellular networks wherein the BSs operate in FD mode, the SI is leaked from the transmit to the receive arrays of the BSs which makes the uplink users corrupted by the SI experience very low spectral efficiency \cite{zfjournal}.

To overcome this limitation, massive MIMO has emerged as a viable solution by employing large number of antennas, i.e., providing enough degree of freedom to sustain enough spatial streams and suppress the SI as well as cellular interference \cite{mimo2}. By increasing the number of antennas in a rich scattering environment, the achievable rate can be enhanced without increasing the bandwidth \cite{mimo3}.

Although combining FD with massive MIMO may circumvent the near-far problem effect, employing a massive number of antennas may require high power consumption.  This is the case for an all-digital systems wherein each RF chain and ADC/DAC is dedicated to a single antenna phase shifter \cite{overview}. To address this shortcoming, using low-resolution ADCs/DACs have been proposed as an efficient solution to reduce the power consumption with a tradeoff in reduction in spectral efficiency \cite{massiveadc,R3,R4,R5}.

In this work, we consider the modeling of FD massive MIMO cellular networks with low-resolution ADCs/DACs and pilot contamination. Pilot contamination occurs when users in a neighboring cell are using a non-orthogonal pilot sequence. The BSs are equipped with a massive number of antennas, operate in FD mode, and have low-resolution ADCs/DACs. The user equipments (UEs) have a single antenna, operate in half-duplex mode, and have full-resolution ADCs/DACs.

System performance is simulated for hexagonal cellular lattice with two tiers. To the best of our knowledge, this is the first work that proposes the modeling and analysis of reverse link (uplink) for FD massive MIMO cellular networks with low-resolution ADC/DAC under pilot contamination. The main contribution of this work is to take into account the many network irregularities and impairments so as to develop a comprehensive analytical framework as well as to show the feasibility of such a system.

Section II presents the network model while analysis of the reverse link is provided by Section III. Numerical results are in Section IV and concluding remarks appear in Section V.

{\bf Notation}. Italic non-bold letters refer to scalars while bold lower and upper case stand for vector and matrix, respectively. We denote subscripts $u$ for uplink and $d$ for downlink.
\section{Network Model}
We consider a macrocellular network where each BS operates in FD mode and is equipped with $N_{\mathsf{a}} \gg 1$ antennas. Each UE operates in half-duplex mode and has a single antenna.
\subsection{Large-Scale Fading}
Each UE is associated with the BS from which it has the highest large-scale channel gain. We denote $K_\ell^u$ and $K_\ell^d$ to be the number of uplink and downlink UEs served by the $\ell$-th BS. The large-scale gain between the $\ell$-th BS and the $k$-th user connected to the $l$-th BS comprises pathloss with exponent $\eta > 2$ and independently and identically distributed (IID) shadowing across the paths, and is defined as
\begin{equation}
G_{\ell,(l,k)} = \frac{L_{\text{ref}}}{r_{\ell,(l,k)}^\eta} \chi_{\ell,(l,k)}  
\end{equation}
where $L_{\text{ref}}$ is the pathloss intercept at a unit distance, $r_{\ell,(l,k)}$ the link distance, and $\chi_{\ell,(l,k)}$ as the shadowing coefficient satisfying $\mathbb{E}[\chi^\delta] < \infty$, where $\delta = 2/\eta$. 

Without loss of generality, we denote the 0-th BS as the focus of interest and we drop its subscript.

\subsection{Small-Scale Fading}
We denote $\boldsymbol{h}_{\ell,(l,k)} \sim \mathcal{N}_{\mathbb{C}}(\bold{0},\boldsymbol{I})$ as the normalized reverse link $N_{\mathsf{a}} \times 1$ small-scale fading between the $k$-th user located in cell $l$ and the BS in cell $\ell$ \cite{massiveforward}. We further denote $\boldsymbol{H}_{\mathsf{SI}} \sim \mathcal{N}_{\mathbb{C}}(\bold{0},\boldsymbol{\mu}_{\text{SI}}^2)$ as the $N_{\mathsf{a}} \times N_{\mathsf{a}}$ SI channel matrix \cite{massiveadc}. We denote by $\boldsymbol{h}_{\ell,(0,k)}$ the channel between the $\ell$-th BS and $k$-th UE in the cell of interest. To the sake of notation, we drop the index 0 for the cell of interest and the fading becomes $\boldsymbol{h}_{\ell,k}$, while the fading the between the $k$-th UE and its served BS in the cell of interest is $\boldsymbol{h}_k$.

\subsection{Full-Duplex and Low Resolution ADC/DAC}
Without loss of generality, we analyze the FD case for low resolution ADC/DAC in single-cell single-user scenario. Then we generalize the analysis for reverse link cellular network.
The received signal $\boldsymbol{y}^u$ at the BS is given by
\begin{equation}
\boldsymbol{y}^u = \sqrt{P_u}\boldsymbol{H}_u\boldsymbol{s}_u + \sqrt{P_d}\boldsymbol{H}_{\mathsf{SI}}\boldsymbol{x}_d + \boldsymbol{n}_u  
\end{equation}
with the downlink unquantized precoded signal $\boldsymbol{x}_d$ given by
\begin{equation}
\boldsymbol{x}_d = \boldsymbol{F}\boldsymbol{s}_d     
\end{equation}
where $\boldsymbol{F}$ is the precoder. As the additive to quantization plus noise model (AQNM) approximates the quantization error \cite{R6,R7,R8,R9}, the received $\boldsymbol{y}^u_q$ and transmitted signals $\boldsymbol{x}_{d,q}$ after the ADC/DAC at the BS can be obtained by
\begin{equation}
\boldsymbol{y}^u_q = \alpha_u \boldsymbol{y}^u + \boldsymbol{q}_u    
\end{equation}
\begin{equation}
   \boldsymbol{x}_{d,q} = \alpha_d \boldsymbol{x}_d + \boldsymbol{q}_d    
\end{equation}
where $\boldsymbol{q}_u$ and $\boldsymbol{q}_d$ are the AQNM for uplink and downlink, while $\alpha = 1-\rho$ and $\rho$ is the inverse of the signal-to-quantization-plus-noise ratio (SQNR), i.e., $\rho \propto 2^{-2b}$. Table \ref{etaparam} defines the values of $\rho$ with respect to the number of bits.
\begin{table}[b]
\renewcommand{\arraystretch}{1}
\caption{$\rho$ for different values of $b$ \cite{massiveadc}.}
\label{etaparam}
\centering
\begin{tabular}{cccccc}
$\boldsymbol{b}$ & 1 & 2 & 3 & 4 & 5\\
\hline
$\boldsymbol{\rho}$ & 0.3634 & 0.1175 & 0.03454 & 0.009497 & 0.002499
\end{tabular}
\end{table} 
We define the AQNM covariance matrices as follows \cite{massiveadc}
\begin{equation}
\boldsymbol{R}_{\boldsymbol{q}_u} = \mathbb{E}[\boldsymbol{q}_u\boldsymbol{q}_u^*] = \alpha_u(1-\alpha_u)\text{diag}\left(P_u\boldsymbol{H}_u\boldsymbol{H}_u^* + \boldsymbol{Q} + \sigma^2 \boldsymbol{I}_{N_{\mathsf{a}}}  \right)    
\end{equation}
\begin{equation}
\boldsymbol{R}_{\boldsymbol{q}_d} = \mathbb{E}[\boldsymbol{q}_d\boldsymbol{q}_d^*] = \alpha_d(1-\alpha_d)\text{diag}\left(\boldsymbol{F}\boldsymbol{F}^* \right)
\end{equation}
where $\boldsymbol{Q}$ is given by
\begin{equation}
\boldsymbol{Q} = P_u\boldsymbol{H}_{\mathsf{SI}}\left( \alpha_d^2\boldsymbol{F}\boldsymbol{F}^* +\boldsymbol{R}_{\boldsymbol{q}_d}  \right)    \boldsymbol{H}_{\mathsf{SI}}^*
\end{equation}
\section{Performance Analysis}
\subsection{Matched Filter Receiver}
Under pilot contamination, a matched filter for user $k$ satisfies $\boldsymbol{w}^{\mathsf{MF}}_k \propto \hat{\boldsymbol{h}}_k$. This entails the following expression as \cite[Eq.~(10.48)]{foundationsmimo} 
\begin{equation}
\begin{split}
\boldsymbol{w}^{\mathsf{MF}}_k =&\sqrt{\frac{\frac{P_k}{P_u}\mathsf{SNR}^u_k}{1+\frac{P_k}{P_u}\mathsf{SNR}^u_k + \sum_{\ell \in \mathcal{C}}\frac{P_{\ell,k}}{P_u}\mathsf{SNR}^u_{\ell,k} } }\\&\times\left( \boldsymbol{h}_k + \sum_{\ell \in \mathcal{C}} \sqrt{\frac{\frac{P_{\ell,k}}{P_u}\mathsf{SNR}^u_{\ell,k}}{\frac{P_k}{P_u}\mathsf{SNR}_k^u}}\boldsymbol{h}_{\ell,k} + \boldsymbol{v}^{'}_k \right)  
\end{split}
\end{equation}
where $\mathsf{SNR}^u_{\ell,k} = G_{\ell,k}P_u/N_0$ is the SNR of the link between the $k$-th uplink in the cell of interest and the $\ell$-th BS and $N_0$ is the noise power. Similarly, $\mathsf{SNR}$ is the $\mathsf{SNR}^u_k$ is the SNR of link between the $k$-th uplink UE and the its serving BS of interest. Note that $\mathcal{C}$ is the set of cells reusing the same pilot dimensions and $P_u$ is the maximum power budget at the UE. The scaling is important to operate the decoder, but immaterial since it equally affects the received signal as well as the noise and interference. With scaling such that $\mathbb{E}\left[\| \boldsymbol{w}_k^{\mathsf{MF}} \|^2\right] = N_{\mathsf{a}}$ and with the entries of $\boldsymbol{v}_k^{\prime}$ having power $1/\left( \frac{P_k}{P_u}\mathsf{SNR}_k^u \right)$. The pilots are assumed to be regular and aligned at every cell.
\begin{corollary}\label{corchannelhardening}
The matched filter receiver $\boldsymbol{w}^{\mathsf{MF}}_k$ has the following properties
\begin{enumerate}
    \item $\mathbb{E}\left[\|\boldsymbol{w}^{\mathsf{MF}}_k \|^2 \right] = N_{\mathsf{a}}$.
    \item $\mathbb{E}\left[\| \boldsymbol{w}^{\mathsf{MF}}_k \|^4 \right] = N_{\mathsf{a}}^2 + N_{\mathsf{a}}$.
    \item $\mathbb{E}\left[ \left| \boldsymbol{w}^{\mathsf{MF}*}_k \boldsymbol{h}_{\ell,\mathrm{k}} \right|^2 \right] = N_{\mathsf{a}}$.
\end{enumerate}
\end{corollary}
\subsection{Channel Hardening}
At the channel estimation stage, we assume the BS operates in half-duplex mode with full-resolution ADCs/DACs. The received signal (sent from the $k$-th uplink UE) at the BS of interest is given by
\begin{equation}
 \boldsymbol{y}^u_k = \sqrt{G_kP_k}\boldsymbol{h}_k + \boldsymbol{v}_k   
\end{equation}

\noindent
Upon data transmission from the users, the BS operates in full-duplex mode with low-resolution ADCs/DACs. Consequently, the BS of interest observes the following received signal vector
\begin{equation}
\begin{split}
\boldsymbol{y}_q ^u=& \alpha_u\sum_\ell \sum_{k=0}^{K_\ell-1} \sqrt{G_{\ell,k}P_{\ell,k}}  \boldsymbol{h}_{\ell,k} s_{\ell,k} +\alpha_u\sqrt{P_{\mathsf{SI}}} \boldsymbol{H}_{\mathsf{SI}}\boldsymbol{q}_d\\&+ \alpha_u\alpha_d\sqrt{P_{\mathsf{SI}}}\sum_{k=0}^{K-1} \boldsymbol{H}_{\mathsf{SI}}\boldsymbol{f}_{k} s_{k}^d + \boldsymbol{q}_u + \alpha_u\boldsymbol{v}     
\end{split}
\end{equation}
By decomposing the channel into estimate and error terms as well as interference into inter-cell, intra-cell, and pilot contamination terms, and after applying the linear receive filter $\boldsymbol{w}^*_k$ at the $k$-th user ($y_{q,k}^u = \boldsymbol{w}^*_k\boldsymbol{y}_q^u$), the received signal at the BS of interest of the $k$-th user is given by (\ref{uplink1}). Note that $\boldsymbol{f}_\text{k}$ is the precoder applied at the BS that is pointing toward the $\text{k}$-th downlink UE. 
\begin{figure*}
\begin{equation}\label{uplink1}
\begin{split}
 y_{q,k}^u =&  \underbrace{\alpha_u\sqrt{G_{k}P_k} \mathbb{E}[\boldsymbol{w}^*_k\boldsymbol{h}_{k}]s_k}_{\textsf{Desired Signal}} + \underbrace{\alpha_u \sqrt{G_{k}P_k}\left( \boldsymbol{w}^*_k\boldsymbol{h}_{k} - \mathbb{E}[\boldsymbol{w}^*_k\boldsymbol{h}_{k}] \right) s_k}_{\textsf{Channel Estimation Error}}+\underbrace{\alpha_u \sum_{\text{k}\neq k} \sqrt{G_{k}P_{\text{k}}}\boldsymbol{w}^*_{k} \boldsymbol{h}_{k}s_{\text{k}}}_{\textsf{Intra-Cell Interference}}+ \underbrace{\alpha_u\sum_{\ell \in \mathcal{C}}\sqrt{G_{\ell,k}P_{\ell,k}}\boldsymbol{w}^*_{k}\boldsymbol{h}_{\ell,k}s_{\ell,k}}_{\textsf{Pilot Contamination}}\\& +\underbrace{ \alpha_u\sum\limits_{\substack{\ell\neq 0\\ 
\ell \notin \mathcal{C} }}\sum_{\text{k}=0}^{K_\ell-1}\sqrt{G_{\ell,k}P_{\ell,\text{k}}}\boldsymbol{w}^*_{k}\boldsymbol{h}_{\ell,k}s_{\ell,\text{k}} + \alpha_u\sum\limits_{
\ell \in \mathcal{C} }\sum\limits_{\substack{\text{k}=0\\\text{k}\neq k}}^{K_\ell-1}\sqrt{G_{\ell,k}P_{\ell,\text{k}}}\boldsymbol{w}^*_{k}\boldsymbol{h}_{\ell,k}s_{\ell,\text{k}}}_{\textsf{Inter-Cell Interference}} +\underbrace{\alpha_u\alpha_d\sqrt{P_{\mathsf{SI}}}\sum_{\text{k}=0}^{K-1}\boldsymbol{w}^*_k\boldsymbol{H}_{\mathsf{SI}}\boldsymbol{f_{\text{k}}}s_{\text{k}}^d}_{\textsf{Self-Interference due to Full-Duplexing}}\\& + \underbrace{\alpha_u\sqrt{P_{\mathsf{SI}}}\boldsymbol{w}^*_k\boldsymbol{H}_{\mathsf{SI}}\boldsymbol{q}_d + \boldsymbol{w}^*_k\boldsymbol{q}_u}_{\textsf{Aggregate AQNM}} +\underbrace{\alpha_u \boldsymbol{w}^*_k\boldsymbol{v}}_{\textsf{Filtered Noise}} 
\end{split}    
\end{equation}
\vspace*{-1cm}
\end{figure*}

\begin{theorem}\label{Theorem4}
For channel hardening and with matched filter receiver, the output SQINR of the $k$-th user is given by (\ref{uplinksqinrhardeningmatched}).
\end{theorem}
\begin{equation}\label{uplinksqinrhardeningmatched}
\overline{\mathsf{sqinr}}_k^{\mathsf{MF}}  = \frac{\alpha_u^2 \left( \frac{P_k}{P_u}\mathsf{SNR}_k^u \right)^2N_{\mathsf{a}}^2}{\left( 1+\frac{P_k}{P_u}\mathsf{SNR}_k^u + \sum_{\ell \in \mathcal{C}}\frac{P_{\ell,k}}{P_u}\mathsf{SNR}_{\ell,k}^u \right) \overline{\mathsf{den}}^\mathsf{MF}}
\end{equation}
where $\mathcal{C}$ is the set of cells reusing the same pilots and $\overline{\mathsf{den}}^\mathsf{MF}$ is given by (\ref{denhardening}). Note that $\frac{P_k}{P_u}$ is the power control coefficient and $\mathsf{INR} = P_{\mathsf{SI}}\mu_{\mathsf{SI}}^2/\sigma^2$ is the interference-to-noise ratio.
\begin{proof}
Using the results of Corollary \ref{corchannelhardening} and \cite[Appendix A]{massiveadc} and after some mathematical manipulations, we retrieve the expression of the SQINR in Theorem \ref{Theorem4}.
\end{proof}
\begin{figure*}
\begin{equation}\label{denhardening}
\begin{split}
\overline{\mathsf{den}}^\mathsf{MF} =& \alpha_u^2N_{\mathsf{a}} \left(1 + \sum_{\ell}\sum_{\text{k}=0}^{K_\ell^u-1} \frac{P_{\ell,\text{k}}}{P_u} \mathsf{SNR}_{\ell,\text{k}}^u \right) + \alpha_u^2N_{\mathsf{a}}^2\frac{\sum_{\ell \in \mathcal{C}}\left( \frac{P_{\ell,k}}{P_u} \mathsf{SNR}_{\ell,k}^u \right)^2}{1 + \frac{P_k}{P_u}\mathsf{SNR}_k^u + \sum_{\ell \in \mathcal{C}}\frac{P_{\ell,k}}{P_u}\mathsf{SNR}_{\ell,k}^u} + \alpha_u^2\alpha_d(1-\alpha_d)K^dN_{\mathsf{a}}^2\mathsf{INR} \\&+ \alpha_u^2\alpha_d^2K^dN_{\mathsf{a}}^2\mathsf{INR} +   N_{\mathsf{a}}\alpha_u(1-\alpha_u)  \left[2 \frac{P_k}{P_u}\mathsf{SNR}_k^u + \sum_{\text{k}\neq k}\frac{P_\text{k}}{P_u}\mathsf{SNR}_{\text{k}}^u + \sum_{\ell\neq 0}\sum_{\text{k}} \frac{P_{\ell,\text{k}}}{P_u}\mathsf{SNR}_{\ell,\text{k}}^u + \alpha_dN_{\mathsf{a}}\mathsf{INR} + 1 \right]      
\end{split}
\end{equation}
\vspace*{-.5cm}
\end{figure*}
With $\overline{\mathsf{sqinr}}_k,~k=0,\ldots K-1$ are locally stable, the evaluation of the gross spectral efficiencies do not require averaging over the fading realizations, but rather it is directly computed as
\begin{equation}
    \frac{\bar{\mathcal{I}}_k}{B} = \log\left( 1 + \overline{\mathsf{sqinr}}_k \right),~k=0,\ldots,K-1
\end{equation}
Considering pilot overhead calls for an aggregation of the reverse and forward spectral efficiencies or a partition of the overhead between the reverse and forward links. Therefore, the effective reverse link spectral efficiency becomes
\begin{equation}
    \frac{\bar{\mathcal{I}}_k^{\mathsf{eff}}}{B} = \left(1 - \beta\frac{N_{\mathsf{p}}}{N_\mathsf{c}}  \right) \log\left( 1 + \overline{\mathsf{sqinr}}_k \right),~k=0,\ldots,K-1
\end{equation}
where $\beta \in [0,1]$ is the fraction of pilot overhead ascribed to the reverse link, $N_{\mathsf{p}}$ and $N_{\mathsf{c}}$ are the number of pilots per cell and the fading coherence tile, respectively.
\begin{corollary}\label{cor2}
To further characterize the spectral efficiency, we derive a new bound using the following formula. Assuming statistical independence between $x$ and $y$, we have
\begin{equation}\label{lemma1}
\mathbb{E}\left[ \log\left(1 + \frac{x}{y} \right) \right] \cong \log\left( 1 + \frac{\mathbb{E}[x]}{\mathbb{E}[y]} \right)    
\end{equation}
\end{corollary}
\begin{lemma}\label{lemmacsi}
Assuming perfect channel state information (CSI), and applying Corollary \ref{cor2}, the output SQINR of the $k$-th uplink user (\ref{uplinksqinrhardeningmatched}) becomes
\begin{equation}
\overline{\mathsf{sqinr}}_k^{\mathsf{MF}}  = \frac{\left( \frac{P_k}{P_u}\mathsf{SNR}_k^u \right)^2(N_{\mathsf{a}}^2+N_{\mathsf{a}})}{\left( 1+\frac{P_k}{P_u}\mathsf{SNR}_k^u  \right) \overline{\mathsf{den}}^\mathsf{MF}}
\end{equation}
While the terms related to the channel estimation error and pilot contamination in (\ref{denhardening}) vanish since the CSI is perfect.
\end{lemma}
\begin{proposition}\label{prop3}
Considering a single-cell multiuser system (without any inter-cell interference) with perfect CSI, Corollary \ref{cor2} entails the results for reverse link in \cite{massiveadc}.
\end{proposition}

\begin{proposition}\label{prop4}
With channel hardening, without full-duplexing ($\boldsymbol{H}_{\mathsf{SI}} = \boldsymbol{0}$), with full-resolution and matched filter receiver, the output SINR of the $k$-th uplink user is given by (\ref{uplinkhardeningsinr}).
\end{proposition}
\begin{figure*}
\begin{equation}\label{uplinkhardeningsinr}
\overline{\mathsf{sinr}}_k^{\mathsf{MF}} = \frac{  \frac{N_{\mathsf{a}}}{1 + \frac{P_k}{P_u}\mathsf{SNR}_k^u + \sum_{\ell \in \mathcal{C}} \frac{P_{\ell,k}}{P_u}\mathsf{SNR}_{\ell,k}^u } \left( \frac{P_k}{P_u}\mathsf{SNR}_k^u \right)^2 }{1 + \sum\limits_{\ell}\sum\limits_{\text{k}=0}^{K_{\ell}^u-1}\frac{P_{\ell,\text{k}}}{P_u}\mathsf{SNR}_{\ell,\text{k}}^u +  \frac{N_{\mathsf{a}}}{1 + \frac{P_k}{P_u}\mathsf{SNR}_k^u + \sum_{\ell \in \mathcal{C}} \frac{P_{\ell,k}}{P_u}\mathsf{SNR}_{\ell,k}^u }\sum\limits_{\ell \in \mathcal{C}} \left( \frac{P_{\ell,k}}{P_u} \mathsf{SNR}_{\ell,k}^u \right)^2 }    
\end{equation}
\vspace*{-.5cm}
\end{figure*}
\begin{remark}
Note that Proposition \ref{prop4} entails the same result for reverse link derived in \cite[Eq.~(10.63)]{foundationsmimo}. 
\end{remark}
\begin{lemma}\label{lemma1}
When the power $\left(\mathsf{SNR}_k^u\right)$ of the $k$-th uplink user goes to infinity, the output SINR (\ref{uplinkhardeningsinr}) converges to $N_{\mathsf{a}}$.
\end{lemma}
This SINR limit, caused by the channel estimation error, is incurred by the users with good conditions, i.e., having small shadowing and short distances from the associated BS. However, this limit is irreducible and cannot be overtaken by a matched filter receiver reliant on channel hardening. 
\begin{lemma}
Neglecting the pilot contamination, the output SINR expression (\ref{uplinkhardeningsinr}) can be reduced to (\ref{uplinksinr}).
\begin{equation}\label{uplinksinr}
 \overline{\mathsf{sinr}}_k^{\mathsf{MF}} \approx \frac{N_{\mathsf{a}} \left( \frac{P_k}{P_u} \mathsf{SNR}_k^u \right)^2 }{ \left(1 + \frac{P_k}{P_u}\mathsf{SNR}_k^u\right)   \left(1 + \sum_\ell \sum_{\text{k}=0}^{K_\ell^u-1} \frac{P_{\ell,\text{k}}}{P_u} \mathsf{SNR}_{\ell,\text{k}}^u \right) }   
\end{equation}
\end{lemma}

\section{Numerical Analysis}
In this section, we present the numerical results along with their discussion. The results are simulated with 10,000 Monte Carlo iterations. In the cell of interest, we first evaluate the SQINR for each UE and then average the SQINRs for all users. Finally, we evaluate the CDF and the spectral efficiency for the average SQINR per cell. Unless otherwise stated, the values of the system parameters are defined in Table \ref{sysparam}.

\begin{table}[t]
\renewcommand{\arraystretch}{.8}
\caption{System Parameters \cite{foundationsmimo,massiveforward}.  }
\label{sysparam}
\centering
\begin{tabular}{rl}
\textbf{Parameter} & \textbf{Value}\\
\hline
Bandwidth & 20 MHz\\
Pathloss Exponent ($\eta$) & 4\\
Shadowing ($\sigma_{\mathsf{dB}}$) & 8 dB \\
Uplink Transmit Power & 200 mW\\
SI Power ($P_{\mathsf{SI}}$) & 40 W\\
SI Channel Power ($\mu_{\mathsf{SI}}$) & 10 dB \\
Thermal Noise Spectral Density & -174 dBm/Hz\\
Noise Figure & 3 dB\\
BS Antennas Gain & 30 dB\\
Number of Antennas ($N_{\mathsf{a}}$) & 100\\
Uplink/Downlink Users per Cell ($K_\ell$) & 10 \\
Number of Pilots per Cell ($N_{\mathsf{p}}$) & 3$K_{\ell}$\\
Fraction of Pilot Overhead ($\beta$) & 0.5 \\
Fading Coherence Tile ($N_{\mathsf{c}}$) & 20,000 (Pedestrians)\\
ADC/DAC resolution & 3 bits
\end{tabular}
\end{table}
\begin{figure}[t]
\centering
\setlength\fheight{5.5cm}
\setlength\fwidth{7.3cm}
\input{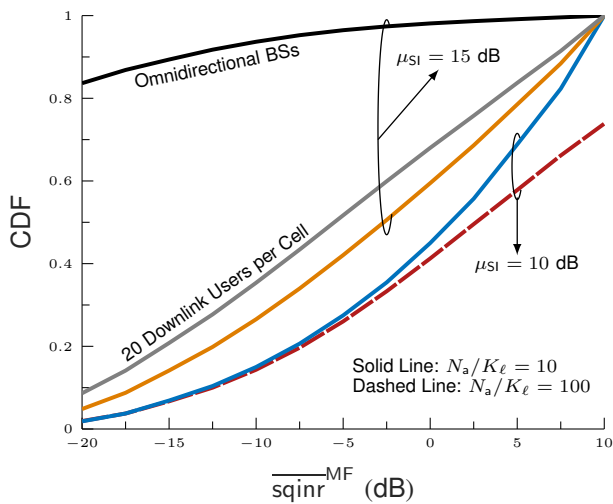}
    \caption{Reverse link results: Effects of the antennas gain, SI channel power and the number of downlink UEs on the CDF of the SQINR. Unless otherwise stated, the number of downlink UEs per cell is 10 users and the antennas array gain is 30 dB. The difference between Red and Blue curves is the the value of the ratio $N_{\mathsf{a}}/K_{\ell}$. The difference between the gray and orange curves is the number of downlink users per cell. The black curve is simulated following the default value but except with 0 dB of antenna gain.}
    \label{fig1}
\end{figure}
Fig.~\ref{fig1} illustrates the CDF of the average reverse link SQINR. When the number of downlink users is large (e.g. 20 users per cell) and hence there is significant SI power, the CDF performance reduces and vice-versa. In addition, by increasing the ratio $N_{\mathsf{a}}/K_\ell$, the user beams become exceedingly narrow, which in turn causes the desired signal to dominate the noise and a large amount of interference to be rejected. When the SI power increases from 10 to 15 dB, the CDF worsens. Furthermore, the antennas' directivity plays an important role in compensating pathloss and suppressing interference. This is confirmed when employing omnidirectional BSs -- the performance completely degrades and vice-versa.
\begin{figure}[t]
\centering
\setlength\fheight{5.5cm}
\setlength\fwidth{7.3cm}
\input{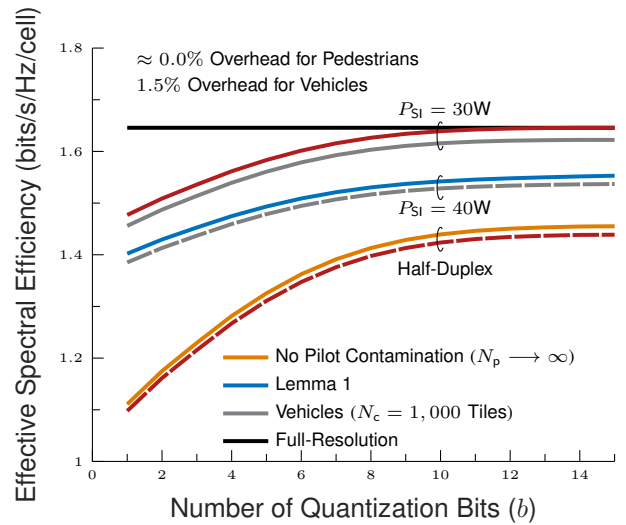}
    \caption{Reverse link results: Effects of SI power, overhead, ADC/DAC resolution, duplexing mode and pilot contamination on the spectral efficiency. The dashed red is simulated for half-duplex and accounting for pilot contamination unlike the solid yellow curve. The dashed gray curve is simulated following the default value as defined in Table \ref{sysparam}. The Hexagonal grid is assumed for this simulation. The solid red curve is considered for pedestrians case unlike the solid gray curve is considered for vehicular scenario.}
    \label{fig2}
\end{figure}
Fig. \ref{fig2} plots the effective spectral efficiency vs. number of ADC/DAC quantization bits ($b$). Spectral efficiency increases with $b$ and converges to a ceiling defined by the full-resolution scheme. The rate decreases when adopting low-resolution ADC/DAC (low $b$). In addition, spectral efficiency noticeably degrades for 40W compared to 30W of SI power. Although FD is vulnerable to loopback SI, it outperforms the half-duplex mode, which is one of the goals of this work. Since the pedestrian model features a large fading coherence tile and hence low overhead, they achieve better effective spectral efficiency compared to the vehicle model (high mobility) that features small fading coherence tile and hence higher overhead.

\section{Conclusion}
In this paper, we propose a unified model for reverse link full-duplex massive MIMO cellular networks with low-resolution ADCs/DACs under pilot contamination. By using matched filter receivers at the base stations, AQNM modeling for ADCs and DACs, and channel hardening, we analyze the SQINR CDF and spectral efficiency for two-tier hexagonal lattice. Using a proper scaling ratio of antennas over the number of users per cell rejects a large amount of interference; however, the SINR is limited by pilot contamination. Simulation results show quantization error and loopback self-interference incur losses; however, this loss is compensated by employing a massive number of antennas. Finally, the proposed system outperforms the half-duplex mode in spectral efficiency, which is part of the goal of this work to show the feasibility of full duplex in cellular networks.

\balance
\bibliographystyle{IEEEtran}
\bibliography{main}
\end{document}